\documentclass[aps,prc,amsfonts,twocolumn,superscriptaddress,showpacs,showkeys,nofootinbib]{revtex4}

\usepackage{color}
\usepackage{graphics,epsfig}
\usepackage{verbatim}

\usepackage{amsmath}
\usepackage{amssymb}
\usepackage{amstext}
\usepackage{mathrsfs}
\usepackage{latexsym}
\usepackage{float}
\usepackage{subfigure}
\usepackage{wasysym}
\usepackage{fancyhdr}

\usepackage{epstopdf}

\newcommand {\be} {\begin{equation}}
\newcommand {\ba} {\begin{eqnarray}}
\newcommand {\ee} {\end{equation}}
\newcommand {\ea} {\end{eqnarray}}

\begin{document}

\newcommand*{\ODU}{Old Dominion University, Norfolk, Virginia 23529}
\affiliation{\ODU}

\title{Observation of a New Type of ``Super''-Symmetry}

\author {S.E.~Kuhn} 
     \thanks{Corresponding author. Email: skuhn@odu.edu}
\affiliation{\ODU}

\date{April 1, 2015}

\begin{abstract}

We report the discovery of an unexpected symmetry that correlates the spin of all elementary particles (integer versus half-integer) with the geographic location of their initial discovery. We find that this correlation is apparently perfect ($R = 1$), with an {\em a priori} probability of $P = 1/65536$ corresponding to a roughly $4.32 \sigma$  deviation from a random distribution.
\end{abstract}

\keywords{Spin, Particle Physics, History of Physics, Discovery}

\maketitle

\section{INTRODUCTION}\label{s1}
The Standard Model of particle physics has been developed in the 60's and 70's of the last century, and
has since withstood numerous experimental tests to very high precision. It describes all (so far) observed
matter as being composed by 12 fundamental fermions with spin 1/2 (in units of $\hbar$) that
interact through their coupling to a set of 5 types of fundamental bosons with spin 0 or 1. All
of these particles have by now been experimentally confirmed, sometimes before and sometimes
significantly after being predicted by or incorporated into the Standard Model. 

It is well known that the Standard Model is not a complete description of nature, for several reasons 
we will not discuss here. A prominent proposed extension of that model entails a new symmetry of nature,
called Supersymmetry, that couples rotations in space-time with rotations in an abstract space of
bosonic versus fermionic degrees of freedom. Among other consequences, Supersymmetry 
predicts the doubling of the aforementioned building blocks of the Standard Model, with a
fermionic  partner for each boson and a bosonic  partner for each fermion. At the time of this
writing, the Large Hadron Collider at CERN is ready to continue its search for such super partners
at the highest available energies. However, no confirmed discovery of any of these additional particles
has been reported up to date.

In this paper, we report the discovery of a different type of ``Supersymmetry'' which applies to the 
already known Standard Model constituents: We observe a perfect correlation between the spin
(integer or half-integer) of a fundamental particle and the location where it was first discovered. In particular, we find that every single boson (the gauge bosons of the electroweak and strong interactions as well as the Higgs boson) was discovered within the geographic space of continental central Europe, while every fermion was  discovered in the Anglo-Saxon hemisphere, encompassing the United States and the United Kingdom. In the following,
we describe our research method and discuss each 
Standard Model elemenatry particle in turn. We conclude with a discussion of the implications 
of our findings.

\section{METHOD}\label{s2}
While modern Physics today is a truly international enterprise, it is fair to say that the historical origin of the scientific revolution underlying it can be traced to the so-called western world of Europe and Great Britain. Roughly during the same time, these countries began colonizing other continents, including the Americas. In particular, the British colonies (and later independent United States) in Northern America can be regarded as the technologically most advanced descendent of this scientific revolution outside of Europe.

Given this history, it seems natural to separate the countries with the most advanced research programs in fundamental Physics, even up to the present day, into two relatively compact geographical areas: On the one hand, the central European continent (designated ``EC'' from here on), and on the other hand, the British Isles and the United States (``BU''). We note that there is a clear border that can be drawn between this two geographic domains, consisting of the Atlantic Ocean, the British Channel and the North Sea, with EC to the East and BU to the West. For the purpose of our study, we set out to find any significant differences between these two domains, as it relates to the discovery of fundamental particles.

In particular, we locate the first discovery or complete description of a new particle (that is now part of the Standard Model) geographically, either by the location of the discovery itself or by the residence of the principal scientists involved. Some care must be taken to properly define what we count as a discovery. In the case of a particle that was discovered {\em after} being predicted or at least conjectured within the framework of the Standard Model, we simply look at the location of the experiment which first announced its discovery (or which, later on, was considered as the crucial step towards acceptance that the particle in question had indeed been observed). However, a few particles were of course discovered before they were predicted - in which case we use the first unambiguous observation, including exclusion of all alternative explanations, of their existence, whether indirectly or directly. We believe that our method is free of any arbitrariness
and therefore our astounding results are not based on any a priori observer bias. We note that, in the
following, we do not distinguish between particles and anti-particles, so that the first observation of
either is counted as the discovery of a specific particle type. It is of note, though, that even the first
anti-particle to be discovered, the positron, was a fermion discovered by ``BU'' physicist C. Anderson.

\section{DATA}\label{s3}
\subsection{Bosons}
The Standard model explains fundamental particle interactions in terms of the exchange of the following elementary gauge bosons: photons ($\gamma$), $W^+$, $W^-$ and $Z^0$ bosons for the electro-weak interaction, and gluons for the strong interaction. In addition, the Higgs boson is required to explain the generation of mass within the Standard Model. We do not consider the graviton, since it has not been discovered yet and cannot be said to be a proper part of the Standard Model in its present form. All five of these bosons have been discovered in the EC region.

{\bf Photon:} The quantization of the electromagnetic radiation was first conjectured by Max Planck, in his attempt to explain the black-body radiation spectrum. The photon hypothesis was concretized by Albert Einstein, who received the Nobel price based on this work. His work was based on experiments by 
Heinrich Hertz and others who studied the effect of electromagnetic radiation of different wave lengths on the emission
of electrons from various metals (photo-electric effect). 
Needless to say, all of these scientists resided in central Europe
when they did this work, mostly in Germany and adjacent countries.

{\bf $W^+$, $W^-$ and $Z^0$ bosons:} 
All three gauge bosons of the weak interaction where discovered in 1983 by the UA1 and UA2 collaborations at the proton-antiproton collider at CERN (headquartered in Geneva, Switzerland).
Even the first indirect evidence for the existence of the $Z^0$ boson, the discovery of neutral
currents, occured at CERN in 1973, with the Gargamelle bubble chamber

{\bf Gluon:} In 1976, M. Gaillard, G. Ross and J. Ellis suggested searching for the gluon via 3-jet events due to gluon bremsstrahlung in $e^+e^-$ collisions. Following this suggestion, the gluon was discovered  in 1979 by TASSO and other experiments using the
 PETRA collider at DESY (Hamburg, Germany).

{\bf Higgs Boson:} The Higgs boson was famously discovered at the Large Hadron Collider (LHC) at CERN, and the discovery announced July 4, 2012. While not all of its properties have been conclusively tested yet, there is little doubt that the discovered boson is at least a close proxy for the Standard Model Higgs. 

\subsection{Fermions}
Within the standard model, there is room for 6 leptons and 6 quarks, all spin-1/2 fermions. All of these have been discovered, as well, and all within either Great Britain or the United States (BU region).

{\bf Electron:} The discovery of the electron is usually credited to J.J. Thompson and his experiments
with cathode ray tubes in 1897 in Great Britain. Further details about the nature of the electron were
unraveled in the 1910 oil drop experiments by American physicist  R.A. Millikan.

{\bf Muon:} The muon was discovered as a constituent of cosmic-ray particle showers in 1936 by the American physicists C.D. Anderson and S. Neddermeyer, and, around the same
time, by J.C. Street and E. C. Stevenson (Harvard Univ.).

{\bf Tauon:} The tau lepton was discovered in 1974--1977 by M. Perl and collaborators at the SPEAR electron-positron
collider at SLAC, Stanford (California).

{\bf Electron neutrino:} C.L. Cowan and F. Reines discovered the electron (anti-)neutrino in 1956, using
the flux from several nuclear reactors in the U.S..

{\bf Muon neutrino:} The muon neutrino was unambiguously identified as a separate neutrino species  by L. Lederman, M. Schwartz and J. Steinberger in 1962, using the Alternating Gradient Synchrotron at the Brookhaven National Laboratory (New York).

{\bf Tau neutrino:} The existence of the tau neutrino was already implied by the discovery of the tauon (see above). Its discovery was announced in July 2000 by the DONUT collaboration working at 
Fermilab  in Batavia (Illinois).

{\bf Up and down quarks:} While quarks never appear as separate entities outside of hadrons like protons and neutrons, 
the first confirmation that such point-like elementary constituents of the proton exist came with the
Deep Inelastic Scattering (DIS) experiments at the then-new Stanford Linear Accelerator Center
(SLAC, California) in the late 1960's, led by J. Friedman, H. Kendall, and R. Taylor. Since their experiment
was mostly sensitive to up and down quarks in the proton, it is credited with the discovery of those specific
two quark types.

{\bf Strange quarks:} The discovery of this quark species is perhaps the most difficult to pin down to a
singular event or place. The first particles containing strange quarks, Kaons, were discovered in
cosmic ray experiments, including those by  G. D. Rochester and C.C. Butler of the University of Manchester (UK) and later ones using cloud chambers on top of Mount Wilson near CalTech (California). 
The correct interpretation of these particles as bound states of strange quarks was first given
with the development of the quark model by M. Gell-Mann and S. Zweig (both U.S. American physicists), which was in turn confirmed
by the same experiments at SLAC described above.

{\bf Charmed quarks:} The first particle containing charmed quarks (and identified as such), the
$J/\psi$, was discovered nearly simultaneously on both coasts of the North American continent,
at SLAC (using the SPEAR ring) and at Brookhaven National Lab. Both discoveries were announced
on November 11, 1974. 

{\bf Bottom and top quarks:} The last two remaining (and heaviest) quark species were both discovered
at Fermilab (Illinois), 18 years apart. The bound state of a bottom and anti-bottom quark, the upsilon, 
was first observed in 1977 using a proton beam and a fixed target. The top quark discovery
required the full energy of the Fermilab Tevatron and occurred in 1995, after a long international
race in both geographic regions (EC and BU).

\section{DISCUSSION AND CONCLUSION}\label{s5}
As demonstrated in the previous section, without a single exception, every boson that is now part of the Standard Model was discovered in the central European continent (EC region), while all of the spin-1/2 fermions were discovered in the BU region (Great Britain and the U.S.). If one assumes an overall set of 17 distinguishable entities (the fundamental particles of the Standard Model), then there are $2^{17}$ possibilities how their discovery could have been (randomly) distributed over two disjoint geographical areas. The perfect correlation we found, on the other hand, would only allow 2 possible cases, depending on which type of particle would be discovered first (boson or lepton). So, without any post hoc bias, we can state that the probability of finding such a perfect correlation as demonstrated here by statistical ``accident'' is 1 in 
$2^{16}$, or $P = 1/65536 = 1.53\times 10^{-5}$. This can be equated to a $4.325\sigma$ deviation from the expectation value. While just shy of a discovery (with the usual criterion of $5\sigma$), this stunning result 
warrants the question whether we can find any mechanism which would explain it.

The answer to this question does not lie in different technologies or overtly different research directions in the two geographical regions we define. Clearly, at any given time, physicists on both sides of the "North Sea border" were keenly aware of the most exciting and promising avenues to search for new particles; several of the actual discoveries described above (including the discovery of the top quark and the Higgs boson) were made after an intense race between several experiments in both regions. 
Both sides of the Atlantic used electron-proton scattering, proton-(anti)proton colliders, and electron-positron colliders to probe physics at the energy frontier. Similar detector technology was widely shared and available
to both sides.

In the absence of a direct causal explanation, a more speculative answer might lie in the different attitudes
and mind sets of the Anglo-Saxon (BU) culture versus those in central Europe, as they have developed from the late nineteenth century on. With the risk of some oversimplification,
the philosophical outlook on life in the U.K. and in the U.S. can be described as more materialistic,
technology/application oriented, and practical. Correspondingly, fermions can reasonably be described as
the building blocks   - the ``bricks'' - of all visible matter around us. It also bears pointing out that, by definition, fermions are more ``individualistic'' - being barred from sharing the same quantum state by the exclusion principle. This correlates with a more individualistic ideal of self sufficiency prevalent particularly in the United States.

On the other hand, central Europe around the beginning of the twentieth century was much more dominated
by a romantic, idealistic view of the world and of Nature, which appears to more naturally conform with
the idea of all-permeating fields as those describing bosons like the photon and the Higgs. In particular,
Europeans were and maybe still are more focused on social interactions and collective goals as opposed to
the accomplishments of the individual. In both positive and negative ways, they have exhibited a greater
tendency to ``bosonic-type condensation''. This mind set was further enhanced among the early giants
of Modern Physics in Europe, who were strongly influenced and fascinated by Eastern Mythology 
(e.g., Werner Heisenberg) or new psychological concepts like the Collective Unconscious (see the exchange between Wolfgang Pauli and C.G. Jung on the idea of ``Synchronicity''). In this context, it surely is no accident that the 
statistical properties of bosons, Bose-Einstein statistics, is named after A. Einstein (a central European)
and the Indian physicist S.N. Bose. On the other hand, fermions follow Fermi-Dirac statistics, after the Italian emigr\'e to the United States, E. Fermi, and British scientist P. Dirac.

As is obvious from the rather speculative nature of these considerations, further research by physicists,
historians of science and other experts is clearly needed to fully explain the perplexing findings 
presented in this paper. However, if there is indeed a deeper mechanism at work, we can make a 
striking prediction which may well be confirmed or falsified within the next months or years: {\bf If}
the LHC discovers a bona fide supersymmetric particle, it will have to be a bosonic partner to
one of the known fermions - a squark or slepton. Only a new accelerator (a linear collider?) built
in Great Britain or the North-American continent will be able to discover the fermionic partners of
the gauge bosons, like gluinos, photinos, Higgsinos, Winos and Zinos.

\section*{Acknowledgments}
I acknowledge valuable discussions with C. Hyde (Old Dominion University) and K. Hartmann (Eastern Virginia Medical School). My research is support by a grant from the U.S. Department of Energy; however, no funds from this grant were used for the research presented here.


\end{document}